\documentclass[12pt]{iopart}
\usepackage{graphicx}

\newcommand{\PrYSO}{Pr$^{3+}$:Y$_2$Si{O$_5$}}

\begin{document}

\title[Cavity enhanced telecom heralded single photons for solid state quantum memories]{Cavity enhanced telecom heralded single photons for spin-wave solid state quantum memories}

\author{Daniel Riel\"ander$^1$, Andreas Lenhard$^1$, Margherita Mazzera$^1$ and Hugues de Riedmatten$^1,^2$}
\address{$^1$ ICFO-Institut de Ciencies Fotoniques, The Barcelona Institute of Science and Technology, 08860 Castelldefels (Barcelona), Spain}
\address{$^2$ ICREA-Instituci\'{o} Catalana de Recerca i Estudis Avan\c{c}ats, 08015 Barcelona, Spain}
\ead{andreas.lenhard@icfo.es}

\begin{abstract}
We report on a source of heralded narrowband ($\approx3$~MHz) single photons compatible with solid-state spin-wave quantum memories based on praseodymium doped crystals. Widely non-degenerate narrow-band photon pairs are generated using cavity enhanced down conversion. One photon from the pair is at telecom wavelengths and serves as heralding signal, while the heralded single photon is at 606 nm, resonant with an optical transition in \PrYSO.  The source offers a heralding efficiency of 28~\% and a generation rate exceeding 2000~pairs/mW in a single-mode. The single photon nature of the heralded field is confirmed by a direct antibunching measurement, with a measured antibunching parameter down to 0.010(4). Moreover, we investigate in detail photon cross- and autocorrelation functions proving non-classical correlations between the two photons. The results presented in this paper represent significant improvement over the state of the art and offer prospects for the demonstration of single photon spin-wave storage in an on-demand solid state quantum memory, heralded by a telecom photon. 
\end{abstract}
\pacs{42.65.Lm}

\noindent{\it Keywords\/}: photon pair source, cavity enhanced SPDC, heralded single photons, narrowband photons, solid state quantum memory

\maketitle

\section{Introduction}
The process of spontaneous parametric down conversion (SPDC) finds widespread application in quantum physics as a source for quantum states of light. In SPDC a pump photon inside an optically nonlinear medium can split spontaneously into a pair of photons (called signal and idler), each of them at lower energy than the pump. The detection of one photon signals the presence of its partner photon. Hence, SPDC can be used as a source of heralded single photons \cite{Grangier1986}. The spectral bandwidth of the SPDC process is typically in the order of hundreds of gigahertz. This results in the generation of broadband photons. However, for many applications in quantum optics, where the photons are interfaced with atomic transitions \cite{Afzelius2015,Bussieres2013}, much smaller line widths are necessary. One possibility to decrease the spectral bandwidth of the emitted photons is to use spectral filters after the crystal \cite{Kaiser2013, Clausen2014, Saglamyurek2014}. However, this requires a very efficient SPDC process to achieve high pair rates and it is technically challenging to build low loss narrowband filtering systems in the MHz range. Another approach is to insert the nonlinear material in a resonator \cite{Ou1999}. In such an optical parametric oscillator (OPO), operated below its oscillation threshold, photons can only be created resonant to the cavity modes. The spectra of the photons resemble the spectrum of the cavity. Additionally, the resonance allows for a spectral enhancement of the creation rate. With this technique, photon sources with line widths in the MHz range could be demonstrated with various designs \cite{Kuklewicz2006,Bao2008,Scholz2009,Wolfgramm2011,Fekete2013, Ahlrichs2016, Luo2015, Foertsch2013,Foertsch2015, Wang2015}. Interfacing such photons effectively with atomic transitions has also been demonstrated \cite{Zhang2011,Rielander2014, Clausen2011, Saglamyurek2011, Schunk2015, Lenhard2015}.

Such sources are of  interest for the development of quantum repeaters \cite{Briegel1998,Sangouard2011}. One interesting scenario is that one photon of the source is stored in a quantum memory while its partner photon is used to distribute entanglement to the neighboring node \cite{Simon2007}. To bridge large distances the second photon should be at a low loss telecommunication wavelength. To that end frequency non-degenerate sources have been developed \cite{Fekete2013, Clausen2014, Saglamyurek2014}. 

Rare earth doped solids cooled to cryogenic temperatures offer suitable atomic transitions to serve photonic quantum memory. Single photons have been stored as collective optical excitations for pre-determined durations $<5 \mu s$ in the excited state of such crystals \cite{Clausen2011,Saglamyurek2011,Rielander2014,Zhou2015,Tang2015} using the atomic frequency comb scheme \cite{Afzelius2009}. Longer storage times and on demand read-out may be reached by transferring the optical excitations to long-lived collective spin-excitations (known as spin-waves) \cite{Afzelius2010,Gundogan2015,Jobez2015}.  Among rare-earth doped materials, praseodymium doped crystals (e.g. \PrYSO) offer exceptional properties for quantum storage. Classical images have been stored for durations on the minute time scale \cite{Heinze2013}, and high storage and retrieval efficiencies have been demonstrated for weak coherent states \cite{Hedges2010} and classical light \cite{Sabooni2013}. Praseodymium ions also offer an electronic structure with three ground state levels allowing spin wave storage. The spin-wave storage of time-bin qubits encoded in weak coherent pulses at single photon level has been shown with praseodymium \cite{Gundogan2015} and europium ions \cite{Jobez2015}. However, the storage of quantum states of light in a spin state of a rare earth is still a remaining challenge and could not be demonstrated, yet. 

A photon pair source compatible with praseodymium doped memories has been demonstrated by our group recently \cite{Fekete2013}, using cavity enhanced down conversion. The source featured a limited heralding efficiency of 6\% and a low detected count rate of around 3 coincidences per mW of pump power, in single mode operation \cite{Rielander2014}. This prevented us to measure directly the single photon character of the emitted light, by measurement of the autocorrelation signal.  While the source was sufficient to demonstrate quantum storage in the excited state of the \PrYSO  crystal \cite{Rielander2014}, its performances were not sufficient for spin-wave storage. A major requirement for the storage in a three level system is a high heralding efficiency, which reduces the number of unnecessary transfer pulses, and an excellent non-classical state at the input of the quantum memory.

In this paper we report on the improved generation of telecom heralded single photons compatible with a Pr quantum memory, using cavity enhanced down-conversion. We report an order of magnitude improvement in detected coincidence count rate, and a significant improvement in heralding efficiency, compared to the state of the art. This allows us to measure directly the single photon nature of the heralded photons through autocorrelation measurements, with measured antibunching parameters down to 0.01. With the help of cross- and unconditioned autocorrelation measurements we demonstrate the violation of the Cauchy-Schwarz inequality. We complete the characterization with a detailed analysis of the non-classical state of our photons for different pump powers and an investigation of the spectral modes formed by the double resonance of the cavity. With these improved performances, our source should now be suitable for spin wave storage in a \PrYSO  quantum memory \cite{Gundogan2015}.

The outline of this report is as follows. We first introduce the experimental setup of the source and discuss the improvements compared to the previous version. It follows a presentation of temporal and spectral parameters of the photons and results for the important figures of merit, the heralding efficiency and the coincidence rates. Afterwards we investigate the non-classical correlations of the photons by cross correlation as well as unconditioned and conditioned auto correlation measurements. We compare the results with theoretical models and check for consistency.

\section{Source}
\begin{figure}[bth]
	\centering
		\includegraphics[width=0.70\textwidth]{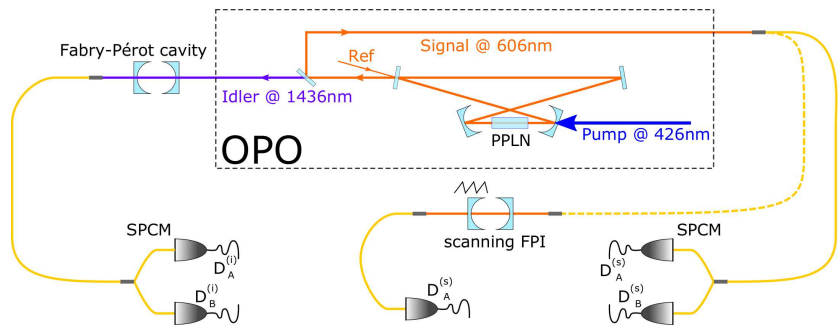}
	\caption{Schematic of the experimental setup. The fiber connections and beam splitters can be changed according to the specific experiment.}
	\label{fig:Setup}
\end{figure}

A sketch of the experimental setup can be found in Fig.~\ref{fig:Setup}.
The photon pairs are generated in a 2~cm long, periodically poled lithium niobate crystal (poling period $\Lambda=16.5$~$\mu$m). We use type-I phase matching where both signal and idler are polarized along the crystal Y-axis. The crystal is temperature stabilized and positioned in the center between the two curved mirrors ($R=100$~mm) of a four mirror bow-tie type ring cavity (free spectral range, FSR=423~MHz). Three mirrors have a highly reflective coating for signal and idler, one of the plane mirrors is used as output coupler. For application with our quantum memory, the wavelength of the signal photons is 606~nm. With the help of a reference laser we actively stabilize our source to the resonance frequency of the QM \cite{Fekete2013}. We use a pump wavelength around 426~nm, resulting in idler photons around 1436~nm in the telecom E-band. In order to ensure that the idler photon is also resonant with the cavity, we use classical light at 1436~nm generated by difference frequency generation between the pump laser at 426~nm and the 606~nm laser. The frequency of this 1436~nm light is actively stabilized to be on resonance by active feed-back on the pump frequency. We use a chopped lock scheme, where we alternate between locking and measurement period at a rate of 30~Hz (see \cite{Fekete2013} for more details). Behind the OPO the photons are split by a dichroic mirror. The idler photons are then sent through an actively stabilized Fabry-Perot filter cavity (linewidth ca. 80~MHz, FSR 17~GHz) before being coupled to a single mode fiber. This additional filter cavity in the idler arm extracts a single mode out of the OPO spectrum and was used in all measurements presented in this paper. The signal photons pass an etalon filter (linewidth 4.25~GHz, FSR 100~GHz) suppressing eventual side clusters and are then coupled to a polarization maintaining single mode fiber. The transmission losses between the cavity output and the detectors are 71(3)~\% for the signal arm, and 35~\% for the idler arm, including 50~\% for the idler filter cavity. We detect the telecom photons with an InGaAs single photon counting module (SPCM, IdQuantique) and the signal photons with a silicon SPCM (Perkin Elmer) with detection efficiencies of 10~\% and 62~\%, respectively. All detectors are fiber coupled. 

In the first design \cite{Fekete2013} of our source, a low transmission out-coupling mirror prevented the photons to escape the cavity with efficiency higher than 30~\%. By reducing the reflectivity of the output mirror to 97~\%, we increased the escape efficiency to ca. 56~\% for the signal photons and 74~\% for idler photons. To calculate the escape efficiency we measure the linewidth and the free spectral range of the cavity to calculate the finesse. As the transmission of the output coupler is known, we can use this value to estimate the internal round trip loss of the cavity (2.4~\% for signal, 1.1~\% for idler). These losses are dominated by absorption in the nonlinear crystal which differs significantly for our signal and idler wavelengths \cite{Schwesyg2010, Waasem2013}. The escape efficiency can be estimated with these internal losses $L_{int}$ and the transmission of the output coupler $T_{oc}$. For a detailed analysis see \ref{apdx:esceff}.

\section{Temporal and spectral characterization}
\begin{figure}[htb]
	\centering
		\includegraphics[width=0.70\textwidth]{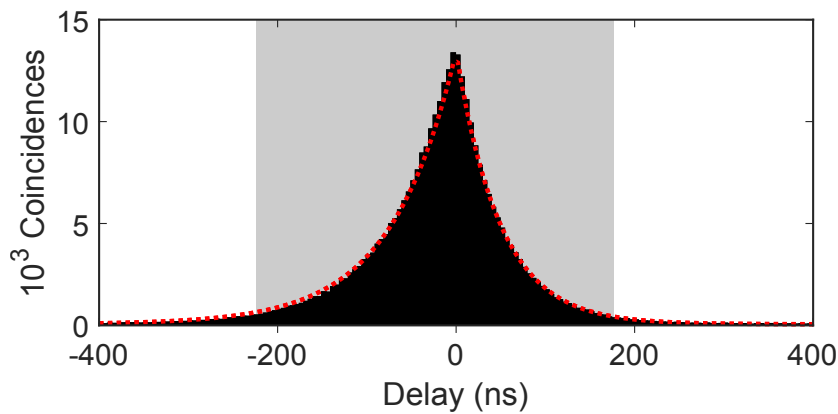}
	\caption{Signal-Idler cross correlation for a pump power of 1~mW. The red dashed line shows a fit to the data as described in the text. The shaded region illustrates the 400~ns integration window. The bin size is 5~ns.}
	\label{fig:CrossCorr}
\end{figure}

As a result of changing the output coupling, the finesse of the resonator was reduced to 184 for idler and 114 for signal, and the bandwidth of the photons became slightly larger than in \cite{Fekete2013}. To investigate the new bandwidth of the photons we use the second order cross-correlation function $G^{(2)}_{s,i}$. We record the detection times of both detectors with fast time stamping electronics (Signadyne). In the post processing we correlate the time delays between the detection times of idler photons (start events) and signal photons (stop events). A resulting histogram is shown in Fig.~\ref{fig:CrossCorr}.

Its temporal shape is a result of the cross correlation of the temporal shapes of both photons. The spectrum of each photon should resemble the Lorentzian shape of the cavity mode spectrum corresponding to an exponential time structure. Hence we see in Fig.~\ref{fig:CrossCorr} the rising exponential of the idler and the falling exponential of the signal wave packet. We fit the histogram with the following function: 
\begin{eqnarray}
G\propto\exp\left[-2\pi\cdot\Delta\nu_s\cdot t\right]\cdot\Theta\left(t\right)+\exp\left[2\pi\cdot\Delta\nu_i\cdot t\right]\cdot\Theta\left(-t\right)+c_0
\end{eqnarray}
which directly results in the bandwidths $\Delta \nu_{s,i}$ of the photons, which is typically 3.7~MHz for signal and 2.3~MHz for idler. We use the heaviside function $\Theta\left(t-t_0\right)$ to distinguish between the idler ($t<t_0$) and signal ($t>t_0$). The smaller value for the idler photons can be explained by a higher cavity finesse due to slightly lower intra-cavity losses. We further define the full width at half maximum (FWHM) of the measured $G^{(2)}_{s,i}$-function as the correlation time $\tau_c$ of the photon pair ($\tau_c=78$~ns). It is connected to the bandwidth of the individual photons via $\tau_c=\frac{\ln{2}}{2\pi\Delta\nu_s}+\frac{\ln{2}}{2\pi\Delta\nu_i}$. The correlation time corresponds to a biphoton bandwidth of $\Delta \nu = \frac{\ln{2}}{\pi\tau_c} = 2.8$~MHz, which is also the linewidth of the heralded single photon.

\begin{figure}[htb]
	\centering
		\includegraphics[width=0.45\textwidth]{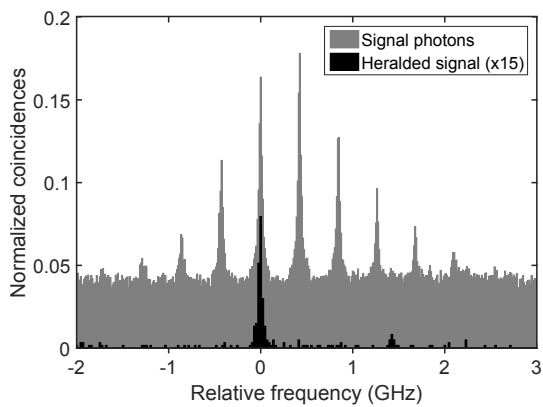}
	\caption{The gray curve shows the spectrum of the signal photons, demonstrating the mode cluster. The black curve shows the spectral distribution of the signal photons heralded by a single mode idler photon.}
	\label{fig:SignalSpectrum}
\end{figure}

The non-degenerate double resonance leads to a clustering effect containing 3 clusters with several modes each \cite{Fekete2013}. We investigate the spectrum of the dominant cluster in the center of the phase matching envelope and show the purity on the longitudinal mode in the idler arm. To this end the signal photons pass a scanning Fabry-Perot cavity (FPI) and the detection events are recorded as well as the scanning trigger. The resulting histogram is shown in Figure~\ref{fig:SignalSpectrum}. The gray curve shows the detection histogram as a function of the cavity length, i.e. relative frequency. Additionally, we recorded the heralding photons which were filtered to a single spectral mode as discussed above. The black curve shows the spectral distribution of the heralded signal photons.
We see that the envelope of the  main cluster contains seven to nine modes and has a width (FWHM) of roughly 1.9 GHz. This is a slightly broader cluster than obtained in \cite{Fekete2013}, due to the lower finesse. However, the heralded detection shows that only one longitudinal mode is transmitted by the idler cavity. The detection events clearly agglomerate at the position of a single longitudinal mode.

\section{Heralding efficiency and coincidence count rate}
\begin{figure}[htb]
	\centering
		\includegraphics[width=0.95\textwidth]{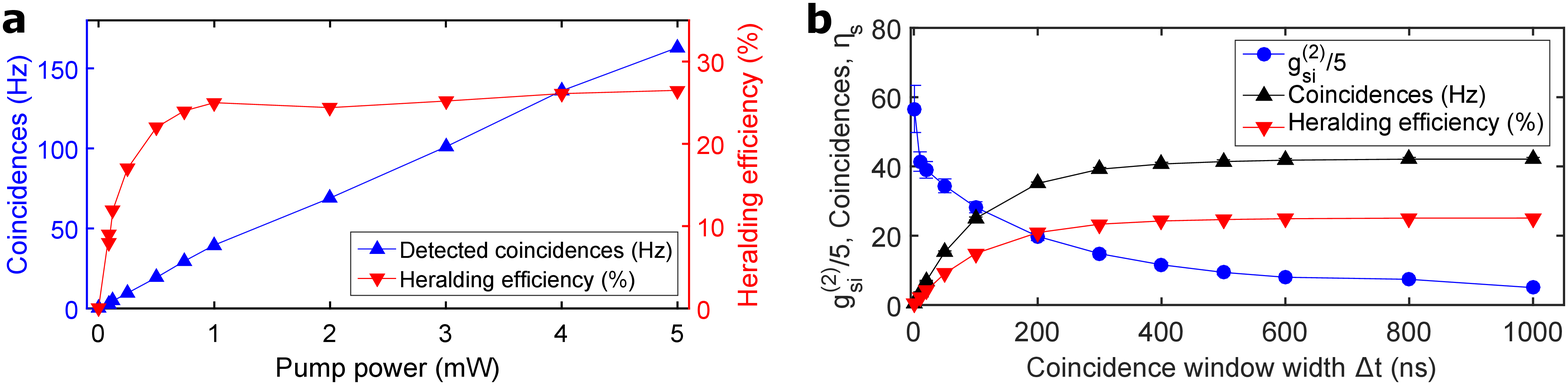}
	\caption{a) Source characterization for different pump powers: Detected coincidences (upward triangles) and heralding efficiency (downward triangles); size of integration window $\Delta \tau=400$~ns. b) Dependence of coincidence rate (upward triangles), heralding efficiency (downward triangles) and cross correlation value (circles) on the choice of the coincidence window width (the $g^{(2)}_{s,i}$ values were divided by a factor of 5 to use the same scale for all measures); pump power 1~mW. Error bars are partly hidden by the markers.}
	\label{fig:Params_vs_Power}
\end{figure}

While the finesse decreased with the higher transmission output coupler, the overall cavity-extracting efficiency of the photons increased. In addition, we increased the transmission of the signal photons between the cavity and the single photon detector. The noise in the idler mode was also reduced  by the use of a new bandpass filter and a new photon detector with a low dark count rate. We use the heralding efficiency $\eta_H=\frac{p_{s,i}}{p_i \eta_{det,s}}$, defined via the coincidence probability, the probability to detect a herald and the detection efficiency of the signal SPCM, as a figure of merit for generating heralded single photons. The improvements resulted in increased heralding efficiencies $\eta_H$ exceeding 28~\% at the signal photon detector, as shown in Fig.~\ref{fig:Params_vs_Power}. This is a significant increase, compared to our previous results in single mode configuration \cite{Rielander2014}. It has been shown that the condition for quantum storage in quantum memories is that $\eta_H > \mu_1$, where $\mu_1$ is the minimum input number of photons to achieve a signal-to-noise ratio of 1 after the memory \cite{Gundogan2015,Jobez2015}. In state of the art demonstrations of solid-state spin-wave quantum memories with weak-coherent states, $\mu_1$ ranges between 0.03 and 0.11 \cite{Kutluer2016,Gundogan2015,Jobez2015}. Our source is therefore promising for the spin-wave storage of single photons in solid-state quantum memories, although current memory demonstrations use weak pulses longer ($>$260~ns) than our single photons. \\

We also measured the coincidence rate for different pump powers. For the detected rate we observe a slope of 34~Hz/mW. Compared to our previous results for single mode operation \cite{Fekete2013} this is an increase by one order of magnitude. Correcting for the measured transmission in the signal and idler arms, we find a creation rate of photon pairs of around $2200 \frac{\mathrm{pairs}}{\mathrm{s}\cdot\mathrm{mW}}$ behind the cavity output mirror for the central mode of the cluster. This gives a spectral brightness at the output of the cavity of approximately $800 \frac{1}{\mathrm{s}\cdot\mathrm{mW}\cdot\mathrm{MHz}}$.  
 
As the coincidences follow an exponential function while the noise is equally distributed in time, the numbers for coincidence rate, heralding efficiency and normalized second-order cross correlation function $g^{(2)}_{s,i}$ will depend on the integration window of the coincidences (the $g^{(2)}_{s,i}$ function will be introduced in the following section). In Fig.~\ref{fig:Params_vs_Power}b we show the dependence of such parameters on the size of the coincidence window, while the window is always centered around the coincidence peak. Obviously coincidence rate and heralding efficiency saturate. For this reason, in this paper when we integrate over a certain window, we use a window width of 400~ns, illustrated in Fig.~\ref{fig:CrossCorr}.

\section{Cross-correlations measurements}
\begin{figure}[htb]
	\centering
		\includegraphics[width=0.75\textwidth]{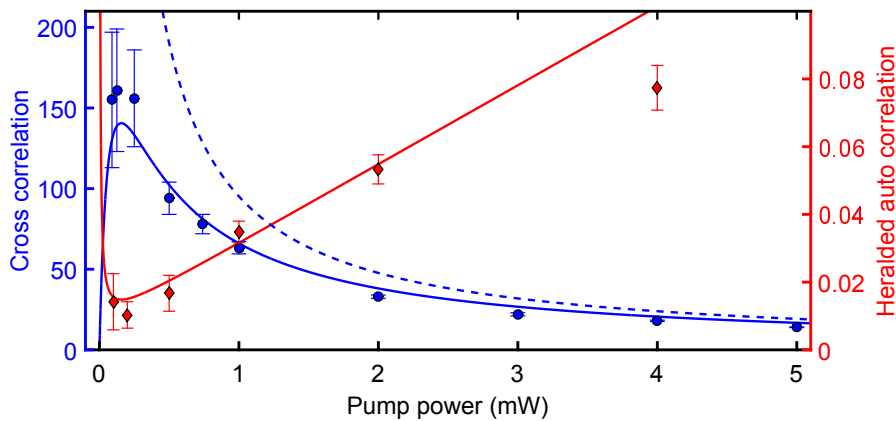}
	\caption{Signal-idler cross correlation $g^{(2)}_{s,i}\left(\Delta\tau\right)$ (blue circles) and heralded signal auto correlation values $g^{(2)}_{i:s,s}$ (red diamonds) for different pump powers. The error bars are calculated assuming square root uncertainty of the count rates. The solid lines show a model describing the data (see text for details). The dashed solid line follows the function $1/\left(p\cdot P\right)$.}
	\label{fig:Correlations}	
\end{figure}

To prove the non-classicality and the single photon character of the signal photons three types of correlation measurements are performed and compared with each other. A first indication for the non-classicality of the generated pairs is given by the normalized cross correlation function $g^{(2)}_{s,i}$. It is defined as $g^{(2)}_{s,i}=\frac{p_{si}}{p_sp_i}$, where $p_{si}$ describes the probability for a coincidence detection of a signal and an idler photon, and $p_{s,i}$ are the detection probabilities for single signal and idler events, respectively. We measure the correlation value with histograms, as the one shown in Fig.~\ref{fig:CrossCorr}. The coincidence rate outside the coincidence peak is a result of accidental coincidences between uncorrelated photons or detector dark counts. We further integrate the coincidence rate in an interval of 400~ns around the correlation peak, shown in Fig.~\ref{fig:CrossCorr}. The ratio of this signal of interest and the background is the $g^{(2)}_{s,i}$-value. The results for different pump powers are illustrated in Fig.~\ref{fig:Correlations}. We measure a maximum value of  $g^{(2)}_{s,i}$ of $161\pm38$, for a pump power of 125~$\mu$W. The value of $g^{(2)}_{s,i}$ is then decreasing for higher pump powers, as expected for a SPDC process. For very low pump powers on the other hand, the rate of detected photons becomes comparable to the dark count rates of the detectors, effectively reducing the $g^{(2)}_{s,i}$-value.
The dependence of $g^{(2)}_{s,i}$ on the pump power, including noise, was theoretically described in \cite{Sekatski2012}. We compare our data with this model, including our detection efficiencies, the dark count rate of our detectors and the inferred photon creation probability $p$. The photon creation probability per time window (400~ns) is calculated from our measured spectral brightness, corrected for the photon bandwidth and escape efficiencies. We find $p\approx 2.2\cdot10^{-3}$~mW$^{-1}$. To describe the data well we use for the dark count probability the average value of both detectors. The model then gives results for $g^{(2)}_{s,i}(0)$-values. For comparison with our measured data we need to modify the values for an integration window of 400~ns. Assuming an exponential decay of the correlation function (with correlation time $\tau_c$) and a background constant in time we find the relation
\begin{equation}
	\frac{g^{(2)}_{s,i}\left(\Delta\tau\right)-1}{g^{(2)}_{s,i}\left(0\right)-1}=\frac{\tau_c}{\Delta\tau}\left[1-\exp{\left(-\frac{\Delta\tau}{\tau_c}\right)}\right]\label{eqn:integrationtime}
\end{equation}
In our case, we find $g^{(2)}_{s,i}\left(\Delta\tau\right)\approx g^{(2)}_{s,i}\left(0\right)/5.16$. The result of the modeling is shown as blue solid line in Fig.~\ref{fig:Correlations} describing well our experimental findings. In the figure we additionally plot the function $1/\left(p\cdot P_{pump}\right)$ (blue dashed line), describing a noiseless cross correlation measurement.

\section{Measurements of unconditional auto-correlation}
\begin{figure}[htb]
	\centering
		\includegraphics[width=0.95\textwidth]{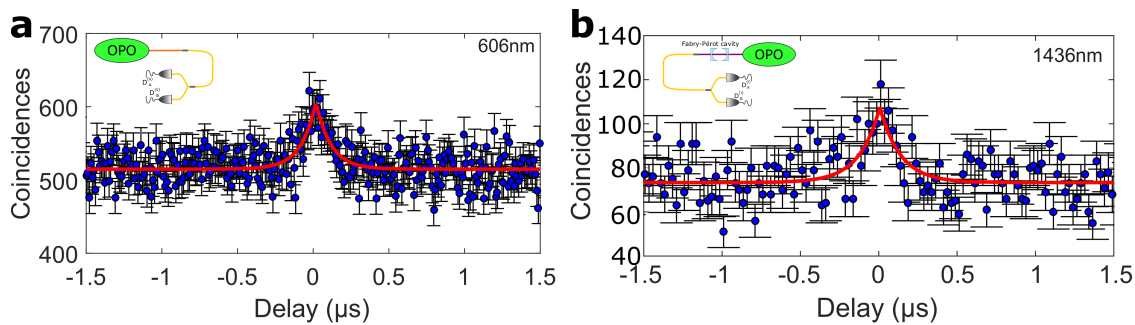}
	\caption{Unconditioned autocorrelations functions for signal (\textbf{a}, pump power 1~mW) and idler (\textbf{b}, pump power 4.3~mW) photons. The solid lines are fits of symmetric exponential functions used to calculate the peak value for zero delay and the peak width.}
	\label{fig:Autocorrelations}
\end{figure}

The signal-idler second-order cross correlation is a quantity easy to measure, but it does not prove the quantumness of the generated light without further assumptions. Hence we performed additional measurements to unambiguously show the quantum character of the correlations. First we measured the unconditional second order autocorrelation functions for signal ($g^{(2)}_{s,s}$) and idler ($g^{(2)}_{i,i}$), shown in Fig.~\ref{fig:Autocorrelations}. To this end we split the particular photons in a 50/50 fiber beam splitter followed by single photon detectors at both outputs. Again we correlate the detection events in the post processing of the time tags, using the data from both the signal detectors only. For an ideal two-mode squeezed state we expect values of $g^{(2)}_{x,x}(0)=2$, which is not the case for our results. For comparison with other measurements we further define the $g^{(2)}_{x,x}$-values via integration over a $\Delta\tau=400$~ns interval, analog to the cross correlation value. This eventually results in measured values of $g^{(2)}_{s,s}(\Delta\tau)=1.10(1)$ and $g^{(2)}_{i,i}(\Delta\tau)=1.32(5)$. Thanks to the use of a low dark count detector this is the first time we observe a bunching peak for the idler photon autocorrelation (see \cite{Rielander2014} for a comparison). From the fits to the histograms we determine the width of the peaks as 138~ns and 184~ns (FWHM; signal and idler, respectively), which is twice the time constant as for the cross correlation, as expected (see \ref{apdx:autocorr}). From the height of the peaks we deduce values of $g^{(2)}_{s,s}(0)=1.18(4)$ for the signal and $g^{(2)}_{i,i}(0)=1.5(2)$ for the idler photons. 

Two reasons can explain the decrease of the measured correlation values: The presence of several spectral modes and noise in the detection process. The measurements, especially for idler, are affected by detector noise. We can estimate this influence with the following basic model that is described in the supplement of \cite{Rielander2014}. The detected count rate $N_{A,B} = S_{A,B} + B_{A,B}$ comprises the signal of interest $S_{A,B}$ and the noise $B_{A,B}$ for both detectors $A,B$. At time delays much larger than the correlation time we expect a coincidence rate proportional to $N_A\cdot N_B$, while it is proportional to $N_AN_B + S_AS_B$ at zero delay, due to the bunching of the SPDC photons. The resulting $g^{(2)}_{x,x}\left(0\right)$-value is then: $g^{(2)}_{x,x,noise}\left(0\right) = 1+\frac{S_AS_B}{N_AN_B}$. We measured dark count rates of $B^{(s)}_A = 30$~Hz and $B^{(s)}_B = 50$~Hz for the signal measurement as well as $B^{(i)}_A = 18$~Hz and $B^{(i)}_B = 192$~Hz for the idler measurements. For the autocorrelation measurement of idler we gated the detectors in phase which the chopper, i.e. the cavity locking cycle, to avoid recording noise when no photons are produced. This is necessary because one of the two telecom detectors has a rather high dark count rate, significantly affecting the measurement outcome. Taking only these detector dark counts into account we find upper limits of $g^{(2)}_{s,s,noise}(0)\leq 1.91$ and $g^{(2)}_{i,i,noise}(0)\leq1.36$. These numbers describe the peak values assuming a single mode and uncorrelated noise. For comparison with our other results we estimate the value that is expected integrating over a 400~ns window (according an equation similar to Eqn.~\ref{eqn:integrationtime}) to be $g^{(2)}_{s,s,noise}(\Delta\tau)\leq 1.57$.

Apart from detection noise we also have to take the effect of multiple modes of the signal photons into account. It has been shown \cite{McNeil1983,Christ2011} that the presence of several modes results in a reduced value following $g^{(2)}_{s,s}(0)=1+\frac{1}{N}$, where $N$ is the number of contributing modes. In the signal branch there is no narrowband filtering installed, which results in the contribution of all modes of the central cluster (see spectrum in Fig.~\ref{fig:SignalSpectrum}). By determining the total intensity as the sum of detected counts in all modes and comparing it to the detections in the central mode, we infer a number of $N=3.9$ effective modes (assuming all modes have the same intensity). This corresponds to $\tilde{g}^{(2)}_{s,s,(MM),noise}(\Delta\tau)=1.15$, close to the measured value.

We can now use the measured data to prove the quantum character of the correlations. The Cauchy-Schwarz inequality $R=\frac{\left(g^{(2)}_{s,i}\right)^2}{g^{(2)}_{s,s}\cdot g^{(2)}_{i,i}}\leq1$ is bound for classical correlations. When we insert the results for cross-correlation ($g^{(2)}_{s,i}\left(\Delta \tau\right)=70(1)$ at 1~mW) and auto-correlation measurements we find $R=2711\pm247$ for an integration window of 400~ns. Hence we violate the classical boundary by more than 10 error margins which is a proof of non-classical correlations.

\section{Heralded narrow-band single-photon source}
\begin{figure}[htb]
	\centering
		\includegraphics[width=0.60\textwidth]{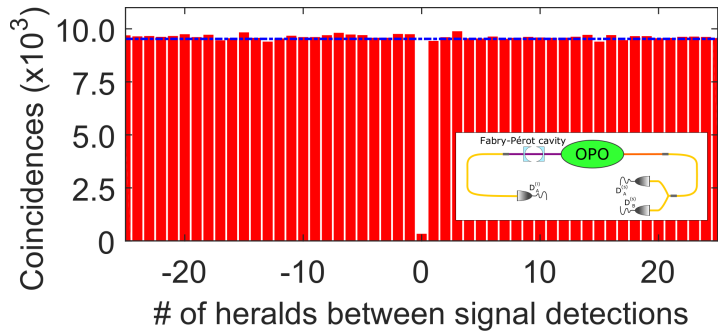}
	\caption{Heralded signal autocorrelation histogram for a pump power of 1~mW. The dashed line illustrates the average height of the outer bins, corresponding to $g^{(2)}_{i:s,s}=1$. For the bin at 0, where triple coincidences are recorded, the height corresponds to $g^{(2)}_{i:s,s}=0.035$.}
	\label{fig:HerAutocorrelation}
\end{figure}

We next check the single photon character of the generated light. This is  done by measuring the autocorrelation conditioned on the detection of a heralding photon. We use a gate window of 400~ns around the detection time of the idler photons (analog to previous measurements) to herald the signal photons. We then use a method introduced by Fasel et al. \cite{Fasel2004} to generate histograms as shown in Fig.~\ref{fig:HerAutocorrelation}. The histogram illustrates the triple coincidences sorted by the number of heralding events between succeeding detections at different signal detectors. The ratio of events in the central bin (bin 0, coincidence of the signal detectors in the same heralding window) divided by the mean value of the outer bins corresponds to the conditioned autocorrelation $g^{(2)}_{i:s,s}=0.035(2)$. This value is considerably below the classical threshold $g^{(2)}_{i:s,s}\geq1$. 

We performed such measurements of the conditioned autocorrelation for various pump powers. The results are shown in Fig.~\ref{fig:Correlations} (red diamonds). For all accessible pump powers the $g^{(2)}_{i:s,s}$-value is well below the classical threshold, with a minimum of 0.010(4). Our source is therefore well suited to generate narrowband heralded single photons with high fidelity. 
The approximately linear dependence of the $g^{(2)}_{i:s,s}$-values on pump power was also observed earlier \cite{Fasel2004}. The inflection of the curve and increase at very low pump powers indicates a regime where the detected photon rate is comparable to the dark count rate. The single photon correlation is then dominated by noise coincidences.

We expect the relation $g^{(2)}_{i:s,s}=\frac{g^{(2)}_{s,s}\cdot g^{(2)}_{i,i}}{g^{(2)}_{s,i}}$ \cite{Chou2004} between conditioned and unconditioned autocorrelation. Hence we can give an estimation with the help of the independently measured values. However, since we condition the detection of the signal photons to the idler, we project the signal photons to a single mode too. Thus we assume for the unconditioned autocorrelation value of the signal photons the theoretical value for a single-mode, corrected for dark counts, and retrieve a prediction of $g^{(2)}_{i:s,s}=\frac{1.57\cdot1.32}{70}=0.03$ for 1~mW of pump power. The red solid line in Fig.~\ref{fig:Correlations} is calculated with this these autocorrelation values while for the cross correlation values the model curve (blue solid line in the same figure) was used. This is in good agreement with experimental results. It should be mentioned that the $g^{(2)}_{i,i}$-value here was measured with a low and a high dark count detector, while for the conditioned autocorrelation only the low dark count detector was used for heralding. Additionally, $g^{(2)}_{i,i}$ was measured only at a high pump power of 4.3~mW. Hence this value is underestimating the quality of the photons.

We summarize the different correlation results, and values derived by different methods in table \ref{tab:g2values} for comparison.
\begin{table}[htb]
\centering
\begin{tabular}{l|c|c|p{3.5cm}|p{2.5cm}}
&$\Delta \tau = 0$&$\Delta \tau = 400$&theoretical&classical\\
& & &prediction&threshold\\
\hline
$g^{(2)}_{i,s}$&335(8)&70(1)&&$\leq\sqrt{g^{(2)}_{s,s} \cdot g^{(2)}_{i,i}}$\\
$g^{(2)}_{s,s}$&1.18(2)&1.10(1)&$\leq 1.15$ ($\Delta\tau=400$)\\
$g^{(2)}_{i,i}$&1.5(2)&1.32(5)&$\leq 1.36$ ($\Delta\tau=0$)\\
R&$63(6)\cdot10^3$&2700(250)&&$\leq1$\\
$g^{(2)}_{i:s,s}$&&0.035(2)&$\leq 0.03$ ($\Delta\tau=400$)&$\geq1$
\end{tabular}
\caption{Second order cross- and auto-correlation and non-classical values $R$ at zero time delay, for an integration window of 400~ns, the theoretical prediction including assumptions for detector dark counts, and the classical boundaries. The pump power is 1 mW for the measurement of all correlation functions, except for $g^{(2)}_{i,i}$ where it is 4.3~mW. }
\label{tab:g2values}
\end{table}

\section{Conclusion}
In summary, we developed a source of telecom-heralded single photons with high heralding efficiency and a high creation rate of narrowband photons compatible with a solid-state quantum memory based on \PrYSO. We proved the single photon character of the emission by direct measurement of the heralded autocorrelation function, with antibunching parameters as low as 0.01. In addition, we reported significant improvements in terms of count rate, quality of correlations and heralding efficiency (up to 28~\%), compared to the state of the art. This source is well suited for advanced experiments with solid state quantum memories. In particular, it meets the requirements to demonstrate single-photon spin-wave storage in a solid-state quantum memory, which would be an important milestone for the use of solid-state quantum memories in quantum networks and quantum repeaters architectures. 

\ack
We acknowledge financial support by the ERC starting grant QuLIMA, by the Spanish Ministry of Economy and Competitiveness (MINECO) and Fondo Europeo de Desarrollo Regional (FEDER) through Grant No. FIS2015-69535-R (MINECO/FEDER), by MINECO Severo Ochoa through Grant No. SEV-2015-0522, by AGAUR via 2014 SGR 1554 and by Fundaci\'o Privada Cellex

\appendix
\section{Estimation of the escape efficiency}\label{apdx:esceff}
In general we have two approaches to estimate the escape efficiency of the photons. We can use the heralding efficiency and calculate backwards to the cavity output or we can use the parameters of the cavity to directly estimate the escape efficiency. Here we want to give a more detailed description of these two ways together with an error estimation.

In figure~\ref{fig:Params_vs_Power} we can read a heralding efficiency of $\eta_h=28$~\%. The heralding efficiency is defined via the probability of detecting a coincidence $p_{si}$, a heralding photon $p_i$ and the detection efficiency $\eta_{det}$ by $\eta_h = \frac{p_{si}}{p_i\cdot\eta_{det}}$. This number can be affected by the measurement accuracy of the detector resulting in $\eta_h=28(1)$~\%. From measurements of the idler spectrum, that look similar to the one shown in Fig.~\ref{fig:SignalSpectrum}, we know that there is a floor of uncorrelated photons present. From the signal to noise ratio, visible in the spectrum, we can estimate that about 10~\% of the detected photons are not correlated with a signal photon. This results in an underestimation of the heralding efficiency, as $p_i$ is corrupted by noise. To estimate the escape efficiency, the heralding efficiency is divided by the transmission losses between cavity and detector, i.e. $\eta_{esc}^{calc}=\frac{\eta_h}{\eta_T}$. For the transmission we found a value of 71(3)~\%. The error accounts for e.g. variations in the fiber coupling and accuracy of the power meter we used, and is found by repetitive measurements. This finally results in $\eta_{esc}^{calc}=39(2)$~\% or $\eta_{esc}^{calc}=44(2)$~\%, if we take the uncorrelated noise into account.

On the other hand, looking at the performance of the cavity we find a finesse of $F=114$ dividing the measured free spectral range by the measured line width. The cavity consists of three high reflecting mirrors (reflectivity $R_{hr}=0.9999$), the output coupling mirror (reflectivity $R_{oc}$) and the crystal. The finesse of the resonator depends on the power ratio recycled after one round trip $\rho$:
\begin{eqnarray}
	F &=& \frac{\pi \rho^{1/4}}{1-\sqrt{\rho}}\\
	\rho &=& R_{hr}^3 \cdot R_{oc} \cdot (1-L_{int})
\end{eqnarray} 
Hence we can infer $\rho$ from the measured finesse. The escape efficiency only depends on the internal losses $L_{int}$ and the output coupling:
\begin{eqnarray}
	\eta_{esc} = \frac{1-R_{oc}}{1-R_{oc}+L_{int}}
\end{eqnarray}
The manufacturer of the mirrors gives a reflectivity of $R_{oc}=0.970(7)$. We solve the above equations numerically and find $L_{int}=2.4(7)$~\% and $\eta_{esc}=56(13)$~\%. The uncertainty on the reflectivity of the out coupling mirror has a big influence on the error bar of the escape efficiency. We also remark that the internal losses include the absorption in the crystal (ca. 0.01~cm$^{-1}$ \cite{Schwesyg2010}; corresponding to 2~\% loss for our case) and losses at the crystal facets (AR coating $R<1$~\%).

Within the error bars we see an agreement between these two methods, resulting in an escape efficiency between 40-50~\% for the signal photons. The escape efficiency of the idler photons is considerably higher due to the lower absorption in the crystal.

\section{Width of autocorrelation function}\label{apdx:autocorr}
The temporal shape of the input wave packet with a Lorentzian spectrum of width $\Delta\nu$ can be described by
\begin{equation}
	f\left(t\right) = \exp\left(-2\pi \cdot \Delta\nu \cdot t\right) \cdot \Theta\left(t\right)
\end{equation}
where $\Theta\left(t\right)$ is the Heaviside step function. The autocorrelation of such a signal can be calculated as follows:
\begin{eqnarray}
	a\left(\tau\right) &=& \int_{-\infty}^{+\infty}{f\left(t\right)\cdot f\left(t+\tau\right) dt}\nonumber\\
	&=& \int_{-\infty}^{+\infty}{\exp\left[-2\pi \cdot \Delta\nu \cdot t\right]\cdot \Theta\left(t\right)} \cdot\exp\left[-2\pi \cdot \Delta\nu \cdot \left(t+\tau\right)\right]\cdot \Theta\left(t+\tau\right) dt\nonumber\\
	&=& \int_{-\infty}^{+\infty}{\exp\left[-2\pi \cdot \Delta\nu \cdot \left(2t+\tau\right)\right] \cdot \Theta\left(t\right) \cdot \Theta\left(t+\tau\right) dt}\nonumber
\end{eqnarray}
To solve the integral we can distinguish two cases for $\tau$:
\begin{eqnarray}
	&&\tau \geq 0 \Longrightarrow \Theta\left(t\right) \cdot \Theta\left(t+\tau\right) = \Theta\left(t\right)\nonumber\\
	a\left(\tau\geq 0\right) &=& \int_{-\infty}^{+\infty}{\exp\left[-2\pi \cdot \Delta\nu \cdot \left(2t+\tau\right)\right] \cdot \Theta\left(t\right) dt}\nonumber\\
	&=& \exp\left[-2\pi \cdot \Delta\nu \cdot \tau\right] \cdot \int_{0}^{\infty}{\exp\left[-2\pi \cdot \Delta\nu \cdot 2t\right] dt}\nonumber\\
	&=& \frac{\exp\left[-2\pi \cdot \Delta\nu \cdot \tau\right]}{4\pi \Delta\nu}
\end{eqnarray}
\begin{eqnarray}
	&&\tau < 0 \Longrightarrow \Theta\left(t\right) \cdot \Theta\left(t+\tau\right) = \Theta\left(t+\tau\right)\nonumber\\
	a\left(\tau < 0\right) &=& \int_{-\infty}^{+\infty}{\exp\left[-2\pi \cdot \Delta\nu \cdot \left(2t+\tau\right)\right] \cdot \Theta\left(t+\tau\right) dt}\nonumber\\
	&=& \exp\left[-2\pi \cdot \Delta\nu \cdot \tau\right] \cdot \int_{-\tau}^{\infty}{\exp\left[-2\pi \cdot \Delta\nu \cdot 2t\right] dt}\nonumber\\
	&=& \frac{\exp\left[2\pi \cdot \Delta\nu \cdot \tau\right]}{4\pi \Delta\nu}
\end{eqnarray}
which means in total:
\begin{eqnarray}
	a\left(\tau\right) &=& \frac{\exp\left(-2\pi \cdot \Delta\nu \cdot \left|\tau\right|\right)}{4\pi \Delta\nu}
\end{eqnarray}
The result for the autocorrelation is a symmetric function where both sides have a decay time given by the spectral width $\Delta\nu$. Hence the full width of the correlation function is twice the width we infer from the cross correlation measurement for signal and idler.\\

\bibliographystyle{iopart-num}
\bibliography{sourcepaper}

\end{document}